\documentclass{article}

\usepackage{arxiv}

\usepackage[utf8]{inputenc} 
\usepackage[T1]{fontenc}    
\usepackage{hyperref}       
\usepackage{url}            
\usepackage{booktabs}       
\usepackage{amsfonts}       
\usepackage{nicefrac}       
\usepackage{microtype}      
\usepackage{lipsum}		
\usepackage{natbib}
\usepackage{doi}
\usepackage{amsmath}
\usepackage{rotating}
\usepackage[graphicx]{realboxes}

\title{Enhanced detection of the presence and severity of COVID-19 from CT scans using lung segmentation}

\author{ \href{https://orcid.org/0000-0003-1274-6750}{\includegraphics[scale=0.06]{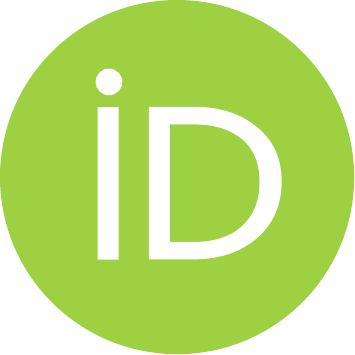}\hspace{1mm}Robert Turnbull} \\
	Melbourne Data Analytics Platform\\
	University of Melbourne\\
	Parkville, Victoria, Australia 3010 \\
	\texttt{robert.turnbull@unimelb.edu.au} \\
}

\renewcommand{\shorttitle}{Enhanced detection of the presence and severity of COVID-19}

\hypersetup{
pdftitle={Enhanced detection},
pdfsubject={cs.AI},
pdfauthor={Robert Turnbull},
pdfkeywords={COVID-19, Computed Tomography, Deep Learning},
}

\begin{document}
\maketitle

\begin{abstract}

Improving automated analysis of medical imaging will provide clinicians more options in providing care for patients. The 2023 AI-enabled Medical Image Analysis Workshop and Covid-19 Diagnosis Competition (AI-MIA-COV19D) provides an opportunity to test and refine machine learning methods for detecting the presence and severity of COVID-19 in patients from CT scans. This paper presents version 2 of Cov3d, a deep learning model submitted in the 2022 competition. The model has been improved through a preprocessing step which segments the lungs in the CT scan and crops the input to this region. It results in a validation macro F1 score for predicting the presence of COVID-19 in the CT scans at 93.2\% which is significantly above the baseline of 74\%. It gives a macro F1 score for predicting the severity of COVID-19 on the validation set for task 2 as 72.8\% which is above the baseline of 38\%.

\end{abstract}


\section{Introduction}

Advances in deep learning offers many opportunities for assisting clinicians in analysis of medical imaging. One possibility for such use is the detection of COVID-19 from computed tomography (CT) scans~\citep{doi:10.1148/radiol.2020200343,kollias2018deep,harmon2020,kollias2020deep,kollias2020transparent}. To encourage research in this area, `AI-enabled Medical Image Analysis Workshop' created a competition as part of the the International Conference on Computer Vision (ICCV) in 2021~\citep{kollias2021mia}. This competition required participants to predict the presence of COVID-19 in a large database of CT scans~\citep{arsenos2022large}. The winning submission achieved a macro F1 score on the test dataset of 90.43\%~\citep{Hou2021}. The competition was run as part of the 2022 European Conference on Computer Vision (ECCV) with an enlarged dataset and an additional task which was to predict the severity of COVID-19 in a subset of CT scans~\citep{kollias2022ai}. Two teams produced winning results in the task to predict the presence of COVID-19 with a macro F1 score of 89.11\%~\citep{Hou2022,Hsu2022}. One of those teams also produced the best submission for the second challenge to predict the severity of COVID-19, achieving a macro F1 score of 51.76\%~\citep{Hou2022}. A third competition is to be held as part of the IEEE International Conference on Acoustics Speech and Signal Processing (ICASSP) in 2023 with an even larger dataset~\citep{kollias2023}. This article presents version 2 of the Cov3d model which achieved a place of `runner-up' in the challenge to detect the presence of COVID-19 in the 2022 competition with a macro F1 score of 87.87\% and a ranking of fourth place in the challenge to predict the severity of COVID-19 with a macro F1 score of 46\%~\citep{turnbull2022}.

\section{The 2023 COV19-CT-DB Database}

The 2023 version of the database includes 1,250 more scans in the training and validation partitions than in the previous year (table \ref{table:partitions}). There were many more non-COVID-19 scans added than positive ones which means that the training dataset contains about 30\% COVID-19 positive scans. The test partition was reduced about 18\% to 4,308.

For a subset of the dataset, two radiologists and two pulmonologists annotated COVID-19 positive CT scans with the severity of the disease in four classes: mild, moderate, severe and critical. The more severity the classification, the more pulmonary parenchymal involvement. These classifications are provided in Microsoft Excel files accompanying the dataset. The number of COVID-19 severity annotations increased substantially to 460 in the training partition and 101 in the validation partition compared to the 2022 version of the database (table \ref{table:severity}). The number of scans in the test partition decreased from 265 to 231.

\begin{figure}
\centering
\includegraphics[width=\linewidth]{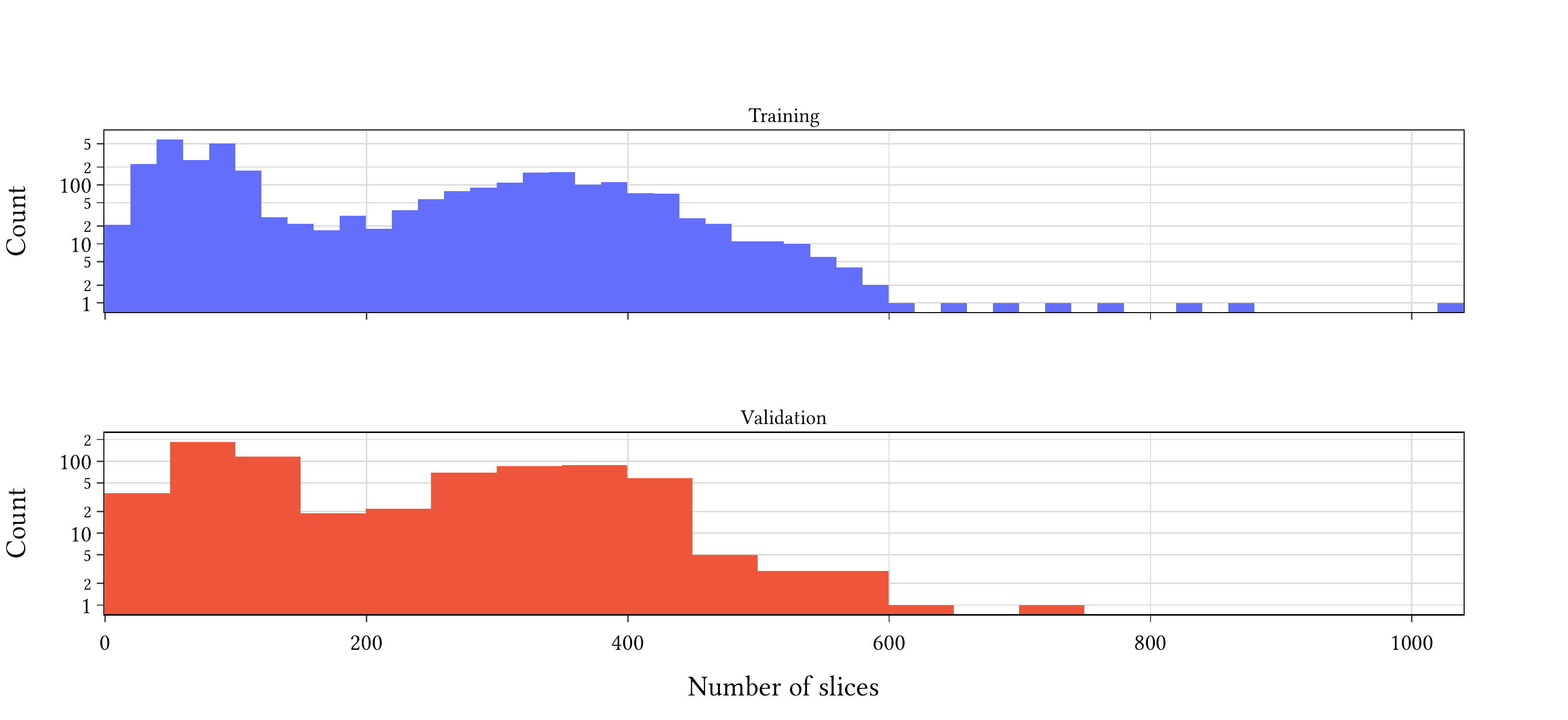}
\caption{The number of transverse plane slices in the scans in the training and validation partitions. The \textit{y}-axis uses a logarithmic scale}
\label{fig:cross-sections}
\end{figure}

The scans are provided as two dimensional JPEG images of cross-sections (slices) in the transverse plane. The images were processed from DICOM files and were clipped to a window of -1150 Hounsfield Units (HU) to 350 HU. Each cross section is of the resolution 512 $\times$ 512 pixels. The number of slices varies considerably (fig. \ref{fig:cross-sections}). Both the training and validation partitions have a bimodal distribution of number of slices, with peaks around 60 and in the mid-300s.

\setlength{\tabcolsep}{4pt}
\begin{table}
\begin{center}
\caption{The number of CT scans in the partitions of the 2023 database with increased number of scans from the 2022 version indicated in parentheses. }
\label{table:partitions}
\begin{tabular}{cccc}
		\toprule
  \noalign{\smallskip}

 COVID-19 & Training & Validation & Test\\
\noalign{\smallskip}
\hline
\noalign{\smallskip}
Positive & 922 (40$\uparrow$) & 225 (10$\uparrow$) & -- \\
Negative & 2,110 (1,000$\uparrow$) & 469 (200$\uparrow$) & -- \\
\noalign{\smallskip}
\hline
\noalign{\smallskip}

Total & 3,032 (1,040$\uparrow$) & 694 (210$\uparrow$) &  4,308 (973$\downarrow$) \\
		\bottomrule
\end{tabular}
\end{center}
\end{table}
\setlength{\tabcolsep}{1.4pt}

\setlength{\tabcolsep}{4pt}
\begin{table}
\begin{center}
\caption{The number of CT scans with severity annotations in the partitions of the database with increased number of scans from the 2022 version indicated in parentheses. These numbers ignore two repeated scans in the annotations.}
\label{table:severity}
\begin{tabular}{ccccc}
\toprule
\noalign{\smallskip}
 Index & Severity & Training & Validation & Test\\
\noalign{\smallskip}
\hline
\noalign{\smallskip}
1 & Mild & 132 (47$\uparrow$) & 31 (9$\uparrow$) & -- \\
2 & Moderate & 123 (61$\uparrow$) & 20 (10$\uparrow$) & -- \\
3 & Severe & 166 (81$\uparrow$) & 45 (21$\uparrow$) & -- \\
4 & Critical & 39 (13$\uparrow$) & 5 (0$\uparrow$) & -- \\
\noalign{\smallskip}
\hline
\noalign{\smallskip}
& Total & 460 (202$\uparrow$) & 101 (40$\uparrow$) &  231 (34$\downarrow$) \\
		\bottomrule
\end{tabular}
\end{center}
\end{table}
\setlength{\tabcolsep}{1.4pt}

\subsection{Cross-Validation}

\begin{figure}
\centering
\includegraphics[width=\linewidth]{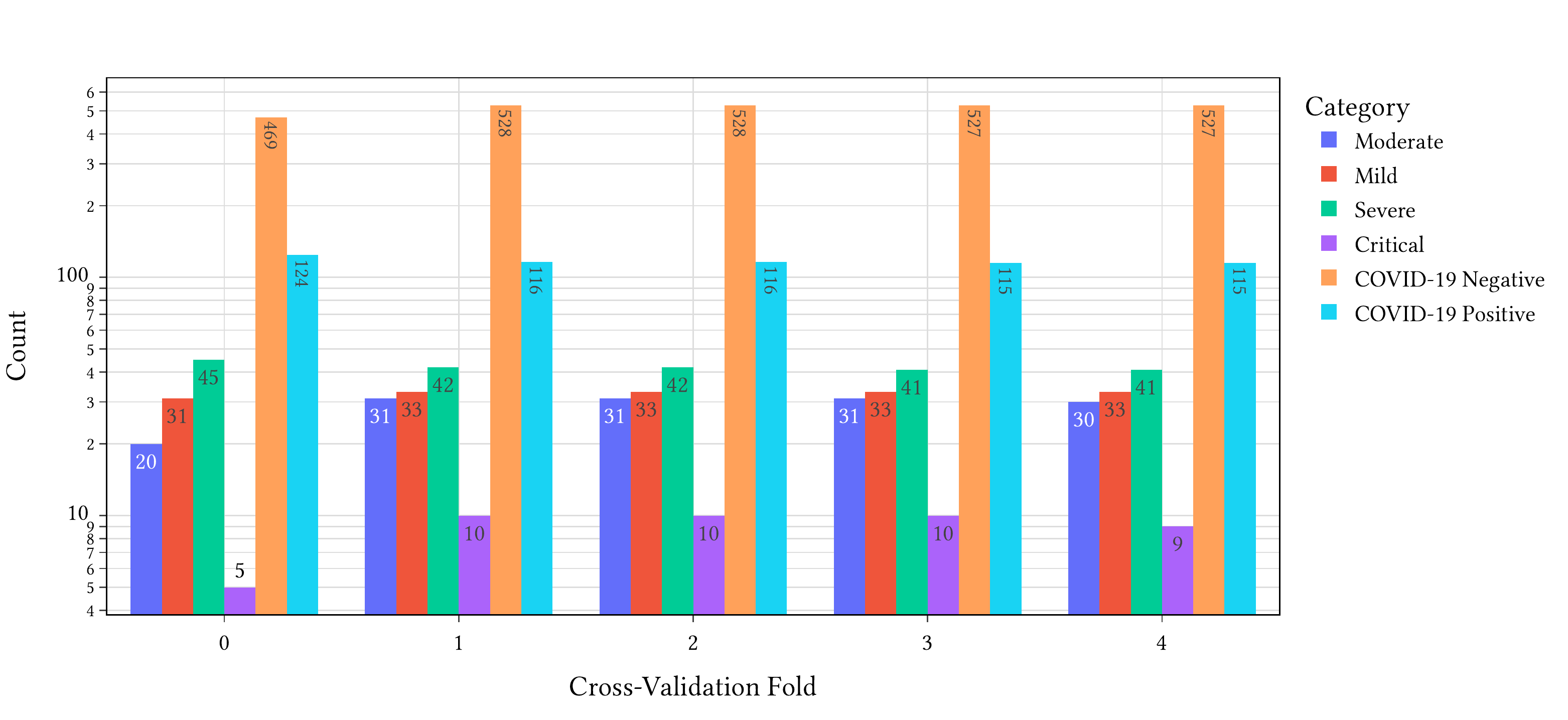}
\caption{The number of scans for each category in the cross-validation folds. The \textit{y}-axis uses a logarithmic scale.}
\label{fig:cross-validation}
\end{figure}

Even with the enlarged database, the number of scans in the validation partition is limited, especially for the different severity categories. The small sample size means that it is difficult to know how well results on the validation partition will reflect on other unseen data. To remedy this we used 5-fold cross validation, as did \cite{Kienzle2022}. We used the existing validation set as one of the folds and labeled fold `zero'. We then split the remaining training data into four evenly split folds whilst keeping a balance of six categories of the scan (i.e. non-COVID, one of the four severity categories or an otherwise unannotated COVID-19 positive scan). This resulted in five fairly evenly balanced training/validation partitions (fig. \ref{fig:cross-validation}).

\section{Methods}

This paper discusses version 2 of Cov3d, a classifier of CT scan images using three dimensional neural networks.\footnote{ \url{https://github.com/rbturnbull/cov3d/}} Cov3d uses TorchApp which is a wrapper for packaging deep learning packages which use PyTorch~\citep{NEURIPS2019_9015} and fastai~\citep{fastai2}.\footnote{TorchApp is available as an alpha release at: \url{https://github.com/rbturnbull/torchapp/}.}

\subsection{Preprocessing}

One challenge of working with the COV19-CT-DB database is the variability of the number of cross-sectional slices (fig. \ref{fig:cross-sections}). Ideally, they will all be a standard size in order to efficiently train the model in batches. Also, the scans with a large number of slices will not be able to fit into the memory for the graphics processing units (GPU) available. That is why the previous version of Cov3d first preprocessed the scan volumes to preset sizes. This usually meant a loss of detail as the volumes were converted to a lower resolution. This can cause a problem if the lungs only constitute a small proportion of the volume since a lot of the detail for detecting the effects of COVID-19 will be lost. To rectify this, first we segmented the lungs in the volume and cropped the volumes to a bounding box containing the voxels identified as part of the lungs. We first attempted to use the `lungmask' neural network software package to perform the segmentation  \citep{hofmanninger2020}. This worked well for slices where the lungs were visible. However, false positive results were produced on slices where no lungs were present. Instead we used an automated procedure of 3D image morphological operations, adapted from \cite{rister2020}. After cropping, the volumes were resized into standard sizes using scikit-image\citep{scikit-image}. The three standard sizes are described in table \ref{table:sizes}.

\setlength{\tabcolsep}{4pt}
\begin{table}
\begin{center}
\caption{The sizes used for preprocessing.}
\label{table:sizes}
\begin{tabular}{cccc}
\toprule
\noalign{\smallskip}
 Description & Depth (Size along longitudinal axis) & Width (Size along sagittal axis) & Height (Size along frontal axis) \\
\noalign{\smallskip}
\hline
\noalign{\smallskip}
Small & 64 & 128 & 128 \\
Medium & 256 & 256 & 176 \\
Large & 320 & 320 & 224 \\

		\bottomrule
  \end{tabular}
\end{center}
\end{table}
\setlength{\tabcolsep}{1.4pt}

\subsection{Loss}

The original Cov3d model used a custom loss function which allowed the model to train simultaneously to predict the presence of COVID-19 and also its severity. This means that the model can be infused with an understanding of the severity of COVID when making predictions about whether or not COVID-19 is present and it allows the severity prediction to be trained on the whole dataset rather than the small subset with the relevant annotations. This combined loss function came through considering the non-COVID-19 scans as being in a fifth severity category. This perspective continued in version 2 of Cov3d but with a different approach.

The loss function for the latest version of Cov3d takes output scores of dimension 5, $z_{c}$, where $c=0$ corresponds to being COVID-19 negative and $c \in {1,2,3,4}$ corresponds to one of the four severity classes. This is converted to a probability distribution with the softmax function:

\begin{equation}
p_{c} = \frac{e^{z_{c}}}{\sum_{j=0}^4 e^{z_{j}}}
\end{equation}

Cov3d then uses the Focal Loss~\citep{focalloss}:

\begin{equation}
\ell_{focal} = - \sum_{c=0}^4 y_{c}(1-p_c)^\gamma \log(p_{c})
\end{equation}
where $gamma$ is a scaling factor and $y_c$ is a one hot encoded vector for the ground truth category, if it is in one of the severity categories or is COVID-19 negative. In the remaining category where the scan is known to be COVID-19 positive but without knowing the severity, the focal loss is calculated as follows:

\begin{equation}
\ell_{focal} = - p_0^\gamma \log(1-p_{0})
\end{equation}

since the probability that the scan is COVID-19 positive is given by $1-p_0$. 

This loss function reduces the contribution of easy examples to classify and forces the network to focus on the more challenging items. However, there is no sense in this loss function that the five categories to be predicted sit somehow along a dimensional axis such that the critical category is closer to the severe category than it is to being mild. For this, we use another component to the loss which is the `Earth Mover's Distance' (EMD) or the Wasserstein distance~\citep{Hou2016}. To calculate this metric, we define a distance between adjacent categories: $[d_{0,1},d_{1,2},d_{2,3},d_{3,4}]$ where $d_{a,b}$ is the distance from moving from category $a$ to category $b$. Then we create a symmetrical distance matrix for all categories:

\begin{equation}
d_{a,b} = \begin{pmatrix}
0 & d_{0,1} & d_{0,1} + d_{1,2} & d_{0,1} + d_{1,2}+ d_{2,3}  & d_{0,1} + d_{1,2}+ d_{2,3} + d_{3,4} \\
d_{0,1} & 0 & d_{1,2} & d_{1,2}+ d_{2,3}  & d_{1,2}+ d_{2,3} + d_{3,4} \\
d_{0,1} + d_{1,2} & d_{1,2} & 0 & d_{2,3}  & d_{2,3} + d_{3,4} \\
d_{0,1} + d_{1,2} + d_{2,3} & d_{1,2} + d_{2,3}& d_{2,3} & 0  & d_{3,4} \\
d_{0,1} + d_{1,2} + d_{2,3} + d_{3,4} & d_{1,2} + d_{2,3} + d_{3,4} & d_{2,3} + d_{3,4} & d_{3,4}  & 0 \\
\end{pmatrix}	
\end{equation}

The EMD loss is then the probability values in each of the five categories multiplied by the distance to move that probability mass into the ground truth category:

\begin{equation}
\ell_{EMD} = \sum_{c=0}^4 y_{c}\sum_{j=0}^4 p_j d_{j,c}
\end{equation}

For the remaining case for COVID-19 positive scans without severity annotations, the distance is zero to all severity categories which are COVID-19 positive and we define a distance $d_{neg-pos}$ which is the distance between the COVID19 negative and positive categories. Thus the EMD loss in this case is:

\begin{equation}
\ell_{EMD} = p_0 d_{neg-pos}
\end{equation}

For the experiments to follow, this distance and distances between adjacent categories $(d_{0,1},d_{1,2},d_{2,3},d_{3,4})$ were all set to 1.

The final loss value is a linear combination of the Focal Loss and EMD Loss:

\begin{equation}
\ell = (1-\lambda)\ell_{focal}+ \lambda \ell_{EMD}
\end{equation}

where $\lambda$ is a hyper-parameter weighting between the two loss components.

\subsection{Neural Network Architecture}

The original Cov3d model used a 3D ResNet neural network architecture~\cite{resnet,Tran2018}. The architecture was adapted to use a single channel for input and to include dropout~\citep{dropout} at each of the four main layers. The final layer was changed to produce the output required. An output of the architecture is shown in table \ref{table:resnet18}.

In addition to the 3D ResNet, a Video Swin Transformer ~\citep{liu2021} was also used with a similar modification to the initial layer and a change to the final layer to the predictions. The `Tiny' version was used for the sake of memory constraints on the GPU.

\setlength{\tabcolsep}{4pt}
\begin{table}
\begin{center}
\caption{The neural network architecture based on the ResNet-18. Square brackets represent convolutional residual blocks. 3D Convolutional layers are represented by the kernel size in the order of depth × width × height and then the number of filters after a comma.}
\label{table:resnet18}
\begin{tabular}{cc}
\hline\noalign{\smallskip}
Layer name             & Description                 \\
\noalign{\smallskip}
\hline
\noalign{\smallskip}
Stem                   & 1×7×7, 64, stride 1×2×2 \\
Layer 1                & $\begin{bmatrix} 3\times3\times3, 64\\ 3\times3\times3, 64 \end{bmatrix}\times2$, Dropout	                  \\
Layer 2                &  $\begin{bmatrix} 3\times3\times3, 128\\ 3\times3\times3, 128 \end{bmatrix}\times2$, Dropout                          \\
Layer 3                &  $\begin{bmatrix} 3\times3\times3, 256\\ 3\times3\times3, 256 \end{bmatrix}\times2$, Dropout                          \\
Layer 4                &  $\begin{bmatrix} 3\times3\times3, 512\\ 3\times3\times3, 512 \end{bmatrix}\times2$, Dropout                          \\
Global Average Pooling & 1 × 1 × 1                  \\
Penultimate Layer      & 512 dimensions, Dropout             \\
Final Layer            & 5 dimensions      \\    
\bottomrule
\end{tabular}
\end{center}
\end{table}
\setlength{\tabcolsep}{1.4pt}

\subsection{Pretraining}

The model was pretrained on a video classification task, the Kinetics-400 dataset~\citep{Kinetics}. In some experiments, the network was further trained on the public STOIC dataset\citep{stoic} of 2,000 CT scans with some being labeled as COVID-19 negative and some as COVID-19 positive in two severity categories. This was trained with the same loss function as above but with fewer severity categories and when used as a pretrained network for the COV19-CT-DB database, the weights for the lower severity category were initially mapped to the `Mild' and `Moderate' categories whilst the higher severity category was mapped to the `Severe' and `Critical' categories.

\subsection{Training Procedure}

The models were trained for 40 or 50 epochs with a batch size of two because of constraints on the GPU memory. Training used the Adam optimization method ~\citep{adam}. The learning rate was scheduled according to Smith's `1cycle' policy~\citep{smith2018} with the maximum learning rate set to $10^{-4}$. The macro F1 scores were calculated using Scikit-learn~\citep{scikit-learn}. The best model weights for each task were saved for inference on the test set.

\subsection{Regularization and Data Augmentation}

To allow for generalization to unseen data, the training dataset was augmented by randomly reflecting the input scans through the sagittal plane. For making inferences on the test set, both the original volume and the reflected volume where given to the network and the final result averaged the probability predictions.

The brightness and contrast of the input volume ($x$, scaled from zero to one) was randomly adjusted by the following transformation:

\begin{equation}
\begin{aligned}
x' &= C (x-0.5) + 0.5 + B \\
\text{where} \\
B &\sim N(0,\sigma_b) \\
C &\sim \text{LogNormal}(0,\sigma_c) \\
\end{aligned}
\end{equation}

Weight decay of $10^{-5}$ was added to the loss function.

\section{Results}

\subsection{Experiments}

Several experiments using different combinations of hyper-parameters were run. The results are found in table \ref{table:results}. Cross validation was not able to be performed on all runs due to computational limitations. The best performing model on the official validation set for task 1 achieved a macro F1 score of 93.2\% which was significantly higher than the baseline of 74\%. The best performing model for task 2 achieved a macro F1 score of 72.8\% which was also much higher than the baseline of 38\%.

\setlength{\tabcolsep}{4pt}
\begin{table}
\begin{center}
\caption{The results of experiments in training the neural network models. If cross validation was performed on all 5 folds then the mean $\pm$ the standard deviation is shown for the macro F1 score for each task.}
\label{table:results}
\Rotatebox{90}{%

\begin{tabular}{cccccccccccccc}
\toprule
Exp. & Arch. & Size   & \shortstack{STOIC\\Pretrain} & Dropout & $\lambda$ & \shortstack{Bright\\ ness \\ $\sigma_b$} & \shortstack{Contrast \\ $\sigma_c$} & Epochs & \shortstack{Validation \\ Task 1 \\ macro F1} & \shortstack{Cross \\ Validation \\ Task 1 \\ macro F1} & \shortstack{Validation \\ Task 2 \\ macro F1} & \shortstack{Cross \\ Validation \\ Task 2 \\ macro F1} \\
\noalign{\smallskip}
\hline
\noalign{\smallskip}
1  & ResNet & Medium & Yes & 0.2 & 0.2 & 0.04 & 0.04 & 50 & 93.2 & 93.5 $\pm$ 0.5 & 60.7 & 64.5 $\pm$ 5.5 \\
2  & ResNet & Medium & Yes & 0.2 & 0.2 & 0.04 & 0.04 & 50 & 93.1 & 93.3 $\pm$ 0.7 & 59.5 & 66.7 $\pm$ 6.6 \\
3  & ResNet & Medium & Yes & 0.2 & 0.1 & 0.04 & 0.04 & 50 & 92.9 & 93.5 $\pm$ 0.5 & 64.7 & 64.7 $\pm$ 4.1 \\
4  & ResNet & Medium & Yes & 0.2 & 0.2 & 0.04 & 0.0  & 50 & 92.8 & 93.4 $\pm$ 0.7 & 57.6 & 61.1 $\pm$ 5.9 \\
5  & ResNet & Large  & Yes & 0.5 & 0.1 & 0.04 & 0.0  & 40 & 92.2 & 93 $\pm$ 1     & 61.9 & 64 $\pm$ 5.9   \\
6  & ResNet & Medium & No  & 0.2 & 0.2 & 0.03 & 0.0  & 50 & 92.1 & 93.4 $\pm$ 1   & 62.0 & 65.3 $\pm$ 6.2 \\
7  & ResNet & Large  & Yes & 0.5 & 0.1 & 0.0  & 0.0  & 40 & 92.1 & 93 $\pm$ 0.7   & 52.5 & 57.3 $\pm$ 6.2 \\
8  & ResNet & Medium & No  & 0.2 & 0.2 & 0.0  & 0.0  & 40 & 92.0 & 93.2 $\pm$ 0.9 & 62.0 & 64.8 $\pm$ 4.4 \\
9  & ResNet & Medium & No  & 0.5 & 0.1 & 0.0  & 0.0  & 40 & 91.7 & 92.6 $\pm$ 0.6 & 61.2 & 60.2 $\pm$ 9.9 \\
10 & ResNet & Medium & No  & 0.5 & 0.5 & 0.0  & 0.0  & 40 & 91.6 & 93 $\pm$ 0.9   & 54.6 & 58.6 $\pm$ 4.2 \\
11 & Swin   & Small  & No  & -   & 0.2 & 0.03 & 0.0  & 50 & 91.5 &                & 68.1 &                \\
12 & ResNet & Small  & No  & 0.5 & 0.1 & 0.0  & 0.0  & 40 & 91.3 &                & 61.4 &                \\
13 & ResNet & Large  & Yes & 0.2 & 0.2 & 0.04 & 0.0  & 50 & 91.2 & 92.8 $\pm$ 1.4 & 68.6 & 63 $\pm$ 8.5   \\
14 & ResNet & Medium & Yes & 0.5 & 0.1 & 0.04 & 0.0  & 40 & 91.2 & 92.8 $\pm$ 1.1 & 51.6 & 61.4 $\pm$ 9.1 \\
15 & ResNet & Small  & No  & 0.5 & 0.5 & 0.0  & 0.0  & 40 & 91.2 &                & 66.3 &                \\
16 & Swin   & Small  & No  & -   & 0.2 & 0.04 & 0.04 & 50 & 91.0 &                & 72.8 &                \\
17 & ResNet & Large  & Yes & 0.2 & 0.2 & 0.04 & 0.0  & 50 & 90.9 & 92.7 $\pm$ 1   & 59.1 & 63.3 $\pm$ 6.7 \\
18 & Swin   & Small  & No  & -   & 0.2 & 0.03 & 0.0  & 40 & 90.9 &                & 70.0 &                \\
19 & Swin   & Small  & No  & -   & 0.1 & 0.0  & 0.0  & 40 & 90.7 &                & 67.0 &                \\
20 & ResNet & Small  & No  & 0.5 & 0.0 & 0.03 & 0.0  & 40 & 90.6 &                & 62.8 &                \\
21 & Swin   & Small  & No  & -   & 0.1 & 0.03 & 0.0  & 40 & 90.6 &                & 67.6 &                \\
22 & ResNet & Small  & No  & 0.5 & 0.9 & 0.0  & 0.0  & 40 & 90.2 &                & 51.0 &               
   \\

\bottomrule
\end{tabular}
}%
\end{center}
\end{table}

\section{Ensembles}
Basic ensemble models were created by averaging the probabilities of the top performing models to predict the presence of COVID-19 for each of the 5 cross-validation folds (\ref{table:ensemble-p}). The best ensemble on the validation set achieved a macro F1 score of 93.4\%. Cross validation showed improvements over the best single model as well.

\setlength{\tabcolsep}{4pt}
\begin{table}
\begin{center}
\caption{Results from ensembles of the five best models to predict the presence of COVID-19. Cross validation was performed on all 5 folds then the mean $\pm$ the standard deviation is shown for the macro F1 score for each task.}
\label{table:ensemble-p}
\begin{tabular}{cccc}
\hline\noalign{\smallskip}
Ensemble Symbol	& Experiments	& \shortstack{Validation \\ Task 1 \\ macro F1} & \shortstack{Cross Validation \\ Task 1 \\ macro F1}                          \\
\noalign{\smallskip}
\hline
\noalign{\smallskip}
EP1 & 1, 2       & 93.2 & 94 $\pm$ 0.6   \\
EP2 & 1, 2, 3     & 93.2 & 93.6 $\pm$ 0.6 \\
EP3 & 1, 2, 3, 4   & 93.4 & 93.8 $\pm$ 0.6 \\
EP4 & 1, 2, 3, 4, 5 & 93.2 & 93.8 $\pm$ 0.8 \\

\bottomrule
\end{tabular}
\end{center}
\end{table}
\setlength{\tabcolsep}{1.4pt}

Similar ensembles were created for the top 5 models for predicting the severity (table \ref{table:ensemble-s}. These did not improve performance over the best single model.

\setlength{\tabcolsep}{4pt}
\begin{table}
\begin{center}
\caption{Results from ensembles of the five best models to predict the severity of COVID-19. Cross validations was not performed.}
\label{table:ensemble-s}
\begin{tabular}{cccc}
\hline\noalign{\smallskip}
Ensemble Symbol	& Experiments	& \shortstack{Validation \\ Task 2 \\ macro F1}              \\
\noalign{\smallskip}
\hline
\noalign{\smallskip}
ES1 & 16, 18 & 68.4 \\
ES2 & 16, 18, 13 & 67.5 \\
ES3 & 16, 18, 13, 11 & 66.5 \\
ES4 & 16, 18, 13, 11, 21 & 66.5 \\

\bottomrule
\end{tabular}
\end{center}
\end{table}
\setlength{\tabcolsep}{1.4pt}

\section{Test Set Prediction Submissions}

The predictions on the test set for five models was submitted for each task. For task 1 to predict the presence of COVID-19, the following models were chosen:
\begin{itemize}
\item Exp. 1: The best performing single model on the validation set for task 1.
\item Exp. 1 Cross Validation: an average of the predictions for the five cross validation folds of experiment 1.
\item Exp. 4 Cross Validation: An average of the predictions for the five cross validation folds for experiment 4 which achieved the highest mean macro F1 score.
\item EP3: The best performing ensemble on the validation set for task 1.
\item EP3 Cross Validation: An average of the predictions for the five cross validation folds for ensemble EP3.

\end{itemize}

For task 2 to predict the severity of COVID-19, the following models were used for the submissions:
\begin{itemize}
\item Exp. 16: The best performing single model on the validation set for task 2.
\item Exp. 18: The second best performing single model on the validation set for task 2.
\item Exp. 13: The third best performing single model on the validation set for task 2.
\item Exp 13 Cross Validation: An average of the predictions for the five cross validation folds for experiment 13.
\item ES1: The best performing ensemble on the validation set for task 2.

\end{itemize}

\textbf{NB. This section will be updated once the results of the competition is announced.}

\section{Conclusion}

This version of Cov3d enhances its ability to detect the presence and severity of COVID-19 from CT scans through preprocessing the volumes to crop the focus on just the lungs and through a different loss function. The results for the two tasks of the challenge both significantly outperform the baseline set for the competition. This new version represents how further refinement of machine learning tools for medical imaging analysis is leading to better results and can open new opportunities for use in clinical settings. 

\section{Acknowledgements}
This research was supported by the University of Melbourne’s Research Computing Services and the Petascale Campus. The work also benefited from computational resources provided by the Faculty of Engineering and Information Technology at the University of Melbourne.

\bibliographystyle{unsrtnat}
\bibliography{references}  






\end{document}